\documentclass[preprint2]{aastex}

\slugcomment{Accepted for publication in \pasp}

\begin{document} 

\title{Hubble Space Telescope Science Metrics}


\author{ Georges Meylan\altaffilmark{1}, Juan P. Madrid, 
         Duccio Macchetto\altaffilmark{1} }
\affil{  Space Telescope Science Institute, 3700 San Martin Drive,
         Baltimore, MD 21218 }
\email{  gmeylan@stsci.edu, madrid@stsci.edu, macchetto@stsci.edu }

\altaffiltext{1}{Affiliated with  the Space Telescope  Division of the
European   Space   Agency,    ESTEC,   Noordwijk,   The   Netherlands}


\begin{abstract}

Since its launch  in April 1990, the Hubble  Space Telescope (HST) has
produced an  increasing flow of scientific results.   The large number
of  refereed  publications  based   on  HST  data  allows  a  detailed
evaluation  of  the  effectiveness  of  this observatory  and  of  its
scientific  programs.  This  paper  presents the  results of  selected
science  metrics related  to paper  counts, citation  counts, citation
history, high-impact papers, and the most productive programs and most
cited papers,  through the  end of 2003.   All these  indicators point
towards the high-quality scientific impact of HST.


\end{abstract}






\section{Introduction} 

The Hubble Space Telescope (HST), orbiting the Earth at an altitude of
about 600 kilometers, is the product of an international collaboration
between the  National Aeronautics and Space  Administration (NASA) and
the  European  Space  Agency  (ESA).   Its  position  high  above  any
atmospheric turbulences and the quality of its instruments provide the
astronomical  community with observation  of excellent  resolution and
sensitivity in  the wavelength  domains of the  ultra-violet, visible,
and near infrared.

HST  has produced  an increasing  flow  of scientific  data since  its
launch by the Space Shuttle Discovery, in April 1990.  After more than
a  decade of  Hubble observations,  the large  number  of publications
based on  HST data provides  a statistically sound basis  to determine
the scientific effectiveness of this observatory.

There  are   numerous  previous  studies  about   science  metrics  in
astronomy :  Abt (1981) studied the behavior  of citation histories
for  papers published  in  1961. Trimble  (1995)  analyzed papers  and
citation  counts   for  papers  based  on  data   collected  at  large
telescopes.  Benn  \& S\'anchez (2001) estimated  the participation of
different facilities in  the most cited papers in  astronomy from 1991
to 1998. Crabtree \& Bryson (2001) examined in detail the productivity
and impact of the Canada-France-Hawaii Telescope (CFHT). 

Improvements in databases for  paper and citation counts have prompted
the  Space  Telescope  Science  Institute  (STScI) to  develop  a  new
standard methodology  to define the  scientific impact of  HST through
quantitative  and  objective metrics.   The  aim  is  twofold: (i)  to
monitor the use of the telescope  in order to improve and maximize its
scientific output through adjustments in the process of the allocation
of observing time, and (ii) to  report to the funding agencies, to the
various governing committees, and  to the astronomical community.  See
Meylan  et al.   (2003) for  a succinct  presentation of  some  of our
metrics results.

There   are  two   straightforward  and   relevant  measures   of  the
effectiveness of a  telescope: the number of refereed  papers based on
data  obtained by  the telescope,  and  the citation  count for  those
papers.  It is  obvious that the full scientific  impact of a facility
may also  be evaluated  through other metrics,  such as the  number of
press releases,  the ``most important'' discoveries, etc.   The aim of
this paper is to show the  results of the first two metrics (paper and
citation counts), both objective  quantities which could be reproduced
and verified by other authors.

The content  of this paper is  the following: Section  2 describes the
way we search  and identify refereed papers using  HST data, Section 3
presents the statistics on the  numbers of papers and their citations,
Section  4 defines  and discusses  the so-called  High  Impact Papers.
Sections 5  and 6 present some  highlights of the  science produced by
HST  through the top  ten most  productive programs  and top  ten most
cited HST  papers.  This  paper is devoted  to HST  publications only;
comparisons  with other  telescopes will  be done  in the  near future
through collaborations with those facilities.


\section{Identification process of Refereed Papers Using HST Data} 

There  have been  various definitions  of what  constitutes  ``a paper
based on HST  data''.  Some of these definitions  require that, of the
total amount of observational data used  in a paper, at least 50~\% of
them  originate from  HST.   Others  do not  include  papers based  on
archival  data.   Since  it  is  impossible to  define  precisely  and
consistently the fraction of HST data  used in a given paper, we adopt
the simplest possible  definition: ``a paper based on  HST data'' is a
paper  benefiting directly from  HST observation.   Such a  paper must
contain  at least:  an  HST image,  an  HST spectrum,  or new  numbers
derived directly  from HST data.   All papers based on  data retrieved
from the HST archives are included.  We take into account papers using
archival data either  for reanalysis or for new  scientific aims. This
broad  definition  has  also   been  adopted  by  other  observatories
(e.g. ESO), but it  has to be clearly stated if the  numbers are to be
used for  comparisons amongst different  facilities, which is  not our
aim in this paper.
 
Most of the  information we use comes directly from  the ADS, the NASA
Astrophysics   Data  System   hosted   in  Cambridge,   USA,  at   the
Harvard-Smithsonian Center for Astrophysics (see Kurtz et~al. 2000).

We run  a boolean  logic query  on the ADS  with the  following search
string: ``HST OR (HUBBLE AND SPACE  AND TELESCOPE) OR WFPC OR WFPC1 OR
WFPC2 OR  (WF AND  PC AND HST)  OR (WIDE  AND FIELD AND  PLANETARY AND
CAMERA) OR FGS OR (FINE AND  GUIDANCE AND SENSORS) OR HSP OR (HIGH AND
SPEED AND PHOTOMETER)  OR FOC OR (FAINT AND OBJECT  AND CAMERA) OR FOS
OR (FAINT AND OBJECT AND SPECTROGRAPH)  OR HRS OR GHRS OR (GODDARD AND
HIGH AND RESOLUTION AND SPECTROGRAPH)  OR STIS OR (SPACE AND TELESCOPE
AND  IMAGING AND  SPECTROGRAPH) OR  NICMOS OR  (NEAR AND  INFRARED AND
CAMERA AND MULTI AND OBJECT  AND SPECTROMETER) OR ACS OR (ADVANCED AND
CAMERA AND SURVEYS)''.

The above query produces a  list of papers, with sometimes wrongs hits
(HST stands  also for Hawaiian  Standard Time!). Each paper  is, then,
downloaded and  read in order to  confirm whether it is  a genuine HST
paper.  Since the ADS allows to query only the abstract of a paper and
not its full  text, hard copies of all  refereed journals are searched
manually by the staff of the STScI Library.

For each identified  HST paper, we search for  the program(s) ID(s) of
the  HST  data  used.   A  link  is  then  established  in  MAST,  the
Multimission  Archive at Space  Telescope, between  the paper  and the
program(s).  There is at least one program ID for each HST paper.  For
each HST  program, the list of  publications that it  has generated is
accessible  on-line to  the  astronomical community  through the  MAST
website \texttt{  http://archive.stsci.edu/hst/search.php} by entering
the proposal-program ID.

Our list of papers recognized  as using HST data is publicly available
on-line and can be accessed  by the astronomical community through the
ADS  at  \texttt{http://adsabs.harvard.edu/},  by activating  the  HST
filter  under  \texttt{select  references  in}. Each  month,  the  ADS
receives from MAST, our updated  list of publications, in an automatic
and electronic way.

It is worth mentioning that the  amount of work required to identify a
paper and link  it to a program is, sometimes,  very onerous.  We have
encountered many stumbling blocks,  often created when authors provide
the wrong  program IDs.   We have even  identified a few  papers which
wrongly claimed to be based on HST data.

In order to  test the completeness of our list  of refereed papers, we
contacted all of  the PIs of programs  in Cycles 4 and 5  for which we
could not find  any refereed publications arising from  their data and
the PIs  confirmed that there  were no additional papers.   However we
expect that a  few papers may have been missed by  our search, but the
number must be very small, certainly less than a few percent.


\section{Paper and Citation Counts Metrics} 

Most of  the HST refereed  papers (about 90  \%) are published  in the
five major  refereed journals, viz., the  Astrophysical Journal (ApJ),
the Astronomical Journal (AJ),  Astronomy and Astrophysics (A\&A), the
Monthly  Notices of the  Royal Astronomical  Society (MNRAS),  and the
Publications of the Astronomical Society of the Pacific (PASP). We, of
course, also count all papers  in the other refereed journals, such as
Nature and Science.  In this paper, we take into account only refereed
publications published by the end of December 2003.


\subsection{Paper Counts per Year}

The number of refereed papers based  on HST data is given in Fig.~1 as
a  function  of the  year  of  publication.   Hubble is  an  extremely
productive telescope: between its launch  in April 1990 and the end of
2003, it  has produced  data directly used  in 4,116  refereed papers.
Following  a strong and  regular increase  of publications  during the
first  eight  years,  the  number  of papers  continued  to  increase,
although at  a slower  pace, during  the last five  years, to  reach a
value of 502 for the year 2003.

The  percentage of  HST papers  published in  the  aforementioned five
major  journals, has  grown from  1\% in  1991 to  7\% in  2003.  Some
special issues of the Astrophysical Journal have been dedicated to HST
papers only:  the first  such issue was  published in March  1991 with
data from  the first  generation of instruments,  while the  last such
issue, devoted  to papers based on  data from the  Advanced Camera for
Surveys  (ACS), appeared  in January  2004 (the  papers in  the latter
issue are not included in this study).

Fourteen years after  its launch, HST continues to  have an increasing
productivity, which can be explained by the regular servicing missions
that   maintain  state-of-the-art   technology   for  its   scientific
instruments  and  spacecraft  systems.   The  rate  of  production  of
refereed  papers  reflects the  increase  in  the ``discovery  space''
provided by the new  scientific instruments deployed in each servicing
mission.

\begin{figure}[!]
\plotone{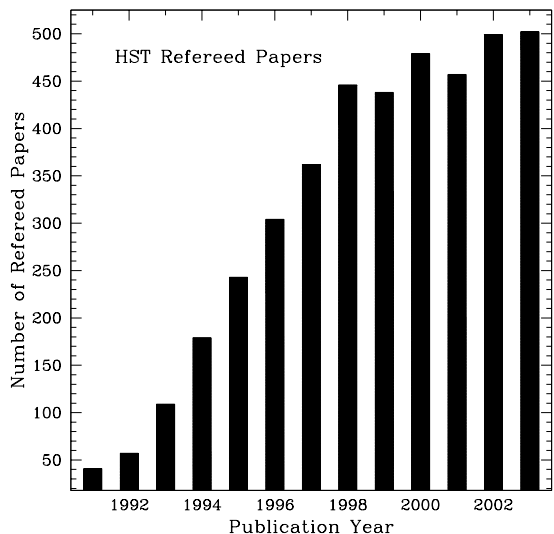}
\caption{Numbers of refereed papers based on HST data as a function of
 the year of publication. \label{fig1}}
\end{figure}


\subsection{Paper Counts per Cycle}

The numbers of refereed papers published per year for all the programs
of a given cycle provide  an interesting metric.  Fig.~2 presents such
information for  Cycles~1 and~2,  Cycles~3 and~4, Cycles~5  and~6, and
Cycles~7 and~8.

In  the upper-left  panel (Cycles~1  and~2), the  initial  increase of
productivity during the first 3-4 years culminates in a peak, which is
followed by a  slow decrease.  Then, the productivity  stabilizes at a
level of about 20-30 papers  per year, 10-12 years after the beginning
of these cycles.

In the upper-right panel, there is an obvious major difference between
Cycles~3 and~4.   The high productivity of Cycle~4  when compared with
Cycle~3 is a  direct consequence of the first  servicing mission which
corrected  the spherical  aberration of  the primary  mirror  with the
installation  of new  instruments.  The  significant increase  in data
quality  triggered a  burst  in publications  using the  significantly
improved performance.

This high level  of scientific output has continued  to increase since
then,  as shown  in Fig.~2  lower-left  panel for  Cycles~5 and~6  and
lower-right  panel for  Cycles~7  and~8.   Each cycle  has  a peak  in
productivity of  about 180  papers reached about  3-4 years  after its
beginning.

\begin{figure}[!h]
\plotone{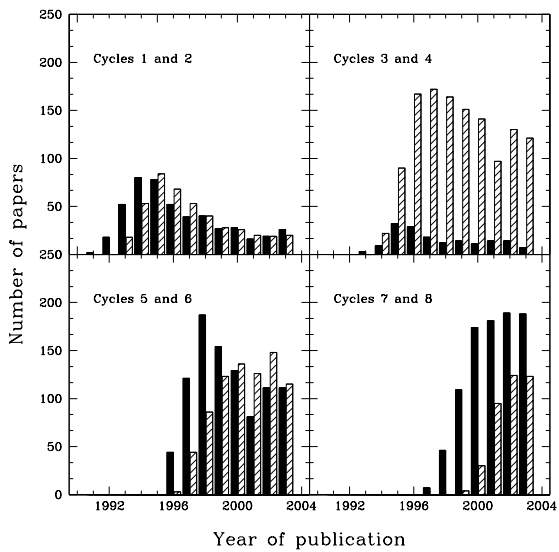}
\caption{ For  Cycles 1  (solid-black histogram) and  2 (cross-hatched
histogram) programs, numbers of refereed papers based on HST data as a
function  of the  year of  publication are  plotted on  the upper-left
panelbox.  Idem for Cycles 3-4, 5-6, and 7-8.
\label{fig2a}}
\end{figure}


\subsection{Paper Counts per Instrument}

Fig.~3 gives the number of refereed papers published per year based on
data from each Hubble instrument. The number in each panel is the time
integral  of the corresponding  curve.  FGS  stands for  Fine Guidance
Sensors,  FOC  for  Faint  Object  Camera, STIS  for  Space  Telescope
Infrared Spectrograph, GHRS  for Goddard High Resolution Spectrograph,
HSP for  High Speed  Photometer, NICMOS for  Near Infrared  Camera and
Multi-Object Spectrograph, FOS for Faint Object Spectrograph, WFPC and
WFPC2 for  the first and second  Wide Field and  Planetary Camera, and
the ACS for Advanced Camera for Surveys.  The FOC, GHRS, HSP, FOS, and
WFPC are  decommissioned instruments, while FGS,  STIS, NICMOS, WFPC2,
and the ACS are  active. A paper may be counted more  than once if the
data it uses come from more than one instrument.

\begin{figure}[!h]
\plotone {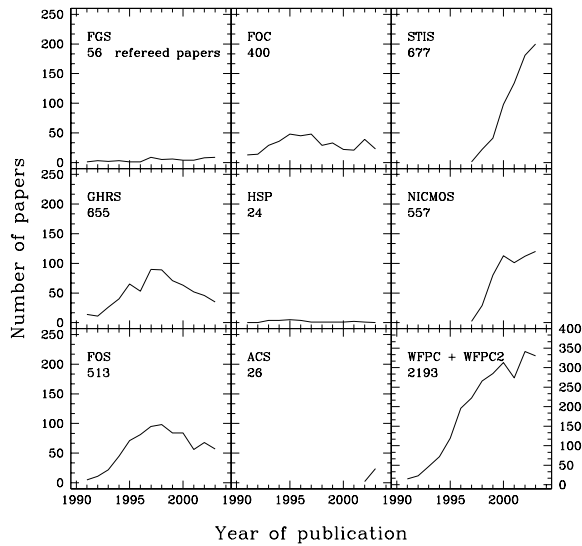}
\caption{Number of refereed papers based  on HST data as a function of
the year  of publication, from launch  to the end of  2003, for active
and  decommissioned HST instruments.   The total  number of  papers in
this figure is 5101, a value larger than 4116, the total number of HST
refereed papers.  This  is due to the fact that  about 1000 papers use
data from more than one instrument and are counted more than once.}
\end{figure}


\subsection{Archives as an Instrument}

By definition, we  consider as an archival HST  paper, any paper which
fulfill simultaneously  the following two conditions: (i)  it is based
on HST  data retrieved  from MAST, the  Multimission Archive  at Space
Telescope, and (ii) none  of its authors is neither a PI  nor a CoI on
any of the HST programs from which these data originate.

Software tools able  to identify such papers have  been developed only
very  recently.  We  currently have  an  estimate of  the fraction  of
archival HST papers only for the  years 2000 through 2003.  Out of the
1870 HST  papers published during  these four years, 654  are archival
HST papers.   This amounts to 34~\%  of the HST papers  (K.  Levay and
MAST team 2004, private communication).


\subsection{Citation Counts per Year}

To estimate the  scientific impact of the refereed  papers based on HST
data, we  obtain from the ADS  the total number of  citations for each
paper in our databases.

The ADS itself is well aware  that its citation counts may suffer from
some (small~?) incompleteness. However the ADS constantly improves its
products  and represents  the most  reliable source  of  citations for
papers published in the major  astronomical journals.  The ADS has two
essential advantages: is it available  on-line and free of any charge.
We use, on a weekly basis,  a script provided by the ADS, which allows
us to access directly the total  number of citations for each paper in
our databases.

\begin{figure}[!h]
\plotone {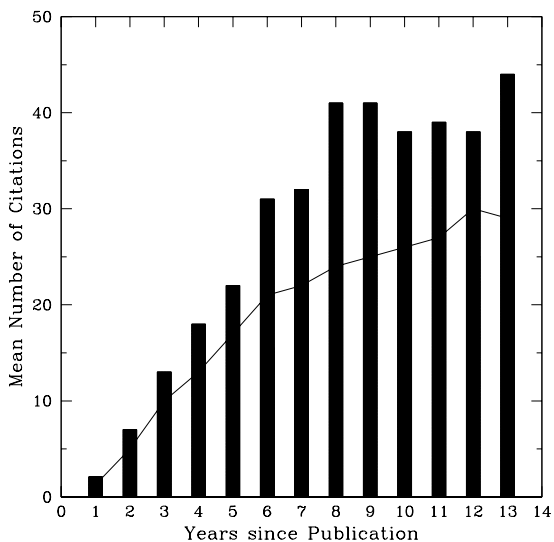}
\caption{Mean number of citations of refereed papers based on HST data
as  a  function of  the  years  since  publication.  The  solid  curve
represents the mean numbers of citations for all astronomy.}
\end{figure}

The Institute for Scientific Information (ISI), based in Philadelphia,
sells  citation counts,  which after  a  few checks,  appear not  more
reliable  than those  from the  ADS. Sandqvist  (2004) presents  a few
examples of large  differences between the ISI and  the ADS because of
errors made by the ISI.  Stevens-Rayburn \& Bouton (2002) also discuss
disparities of citation counts for the two providers.


For Fig.~4, we  consider only papers in the  five major journals (ApJ,
AJ, A\&A, MNRAS,  and PASP), since other journals,  such as Nature and
Science, would  bias our  statistics with their  numerous highly-cited
articles not  related to astronomy.  The histogram  in Fig.~4 displays
the  mean  total number  of  citations of  refereed  HST  papers as  a
function of  years since publication. For the  older papers, published
between 7  and 12  years ago,  the mean total  number of  citations is
about 40 per paper, and is smaller for more recently published papers.
The  segmented line in  this figure  shows the  mean total  numbers of
citations, for  all astronomy papers in the  aforementioned five major
journals.  The refereed HST papers have an average number of citations
per  paper  larger (by  at  least 25\%)  than  the  average number  of
citations of all the astronomical papers.

Fig.~5 shows  that, a few years  after publication, only  about 2\% of
the refereed HST  papers have no citations, whereas  about one quarter
of all refereed  papers in astronomy have no  recorded citation in the
ADS.

These two figures show that  the refereed HST papers have a scientific
impact significantly above average.

\begin{figure}[!h]
\plotone{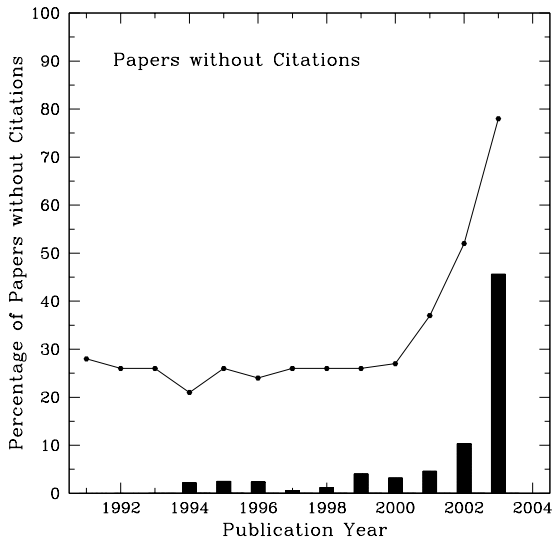}
\caption{Percentage  of refereed  HST  papers without  citations as  a
function of  the year of  publication, as solid-black  histogram.  The
solid  curve  represents the  percentage  of  refereed papers  without
citations for all astronomy.}
\end{figure}


\subsection{Citation History}

The citation counts allow the study of the evolution of citation rates
of  papers as  a function  of the  years following  publication.  Such
metrics  allow to  answer the  following  questions: How  fast is  the
growth in  citation rate~? Is  there a maximum citation  rate~? After
how many  years is this maximum  reached~? How fast is  the decline in
citation rate~?

We have  obtained the individual  citations of all the  4,116 refereed
HST papers,  courtesy of  the ADS (Kurtz,  2003a).  This amounts  to a
total  of  64,141 citations.   Fig.~6  presents  the average  citation
counts  per  paper  as  a  function  of  the  number  of  years  since
publication.  The continuous  line is the average for  the HST papers,
while  the  dashed line  is  the average  of  all  refereed papers  in
astronomy.

In both cases after a sharp rise, the peak of the citation rate occurs
approximately two  years after publication.  HST  refereed papers peak
at an average of 5.9 citations/paper/year, while the peak reaches only
3.2  citations/paper/year  for   all  refereed  papers  in  astronomy.
Thereafter,  the  citation  rate  decreases  linearly  to  be  about  2
citations/paper/year ten years after publication.

Crabtree  \& Bryson  (2001,  2002) generated  similar  curves for  the
papers   produced  from   observations  by   the  Canada-France-Hawaii
Telescope (CFHT) and the UK Infra-Red Telescope (UKIRT).  Their curves
peak at  about 4.0  - 4.5 citations/paper/year,  and display  the same
general  shape; a  sharp  rise,  a maximum  reached  after two  years,
followed  by  a slow  decline.   Abt  (1981)  showed that  for  papers
published  in  1961,  citations  reach  a  maximum  five  years  after
publication.   The delay  between  publication date  and  the peak  of
citations has  shortened since that time  to an average  of two years.
This may  be a direct consequence  of the strong  increase, during the
last decade, in the spread of information through the World Wide Web.

\begin{figure}[!h]
\plotone {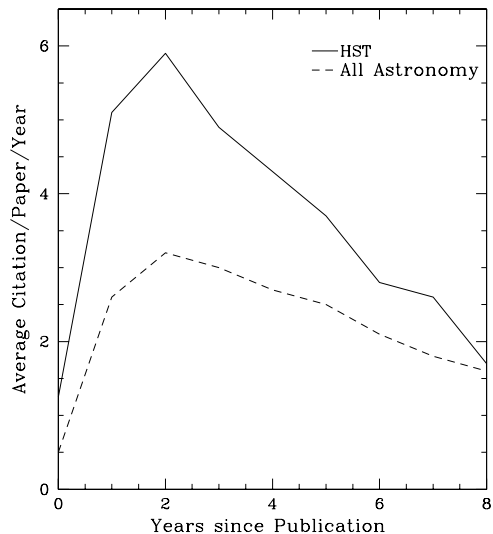}
\caption{Citation history: average citations  rate per paper per year.
The  continuous curve  represents  the average  for  the refereed  HST
papers,  while  the  dashed  curve  represents  the  average  for  all
astronomy papers.}
\end{figure}


\subsection{ADS 2000-2001 reads of papers}

For each  paper, each  access through the  ADS generates a  log entry,
which is called  a ``read''.  An independent and  new way of measuring
the  readership of a  paper is  the number  of ``reads''  it generates
(Kurtz, 2003b).  Michael Kurtz of the ADS kindly provided us with data
related to this new metric.

During the two years 2000 and  2001, the ADS recorded about 12 million
reads of all astronomy papers, out of which 10 million were related to
refereed  papers.  At  that time,  the then  3113 HST  refereed papers
accumulated  533,362 reads  during these  two years,  corresponding to
5.3\% of all reads of astronomy papers.
 

\section{High Impact Papers}

There  is more  than one  way  to identify  papers which  have a  high
scientific impact.  An interesting metric is given by the papers which
have  the largest  numbers  of citations:  called  High Impact  Papers
(HIPs) by  the ISI (see http://www.isinet.com/rsg/hip).   We adopt the
ISI definition: a paper  is a HIP if it belongs to  the 200 most cited
refereed papers published in a given year.  The ADS has the capability
to sort  all papers by citation  counts and publication  dates.  It is
straightforward, from the ADS, to  obtain the top 200 papers published
in a given year. 

We identify the HST refereed  papers which have enough citations to be
among  the  HIPs published  in  a  given  year.  Fig.~7  provides  the
percentage  of HIPs based  on HST  data as  a function  of publication
year. After  a slow start in 1991,  1992, and 1993, the  effect of the
successful deployment  of COSTAR and  WFPC2 on the impact  of refereed
papers  is obvious  between 1993  and  1994.  Since  1994, Hubble  has
consistently generated about 8\% of all HIPs.

We have extracted the High Impact Papers for the following four years:
1998, 1999, 2000,  and 2001, and studied in  greater details these 800
papers.  Namely,  for each paper we  accessed the full  text using the
ADS,  read the  paper and  decide whether  it is  an  observational or
theory paper. Hereafter we consider only observational papers. If, for
a paper, there is more  than one telescope providing the observations,
the weight or percentage of the contribution coming from each facility
is roughly estimated  (e.g., 50 \% HST, 25\%  Keck, and 25\% Chandra),
and the citations of this paper  are attributed to the facilities as a
function  of these  percentages.  In  this  way, the  total number  of
citations related to the 200 HIPs of a given year is distributed among
all telescopes/facilities that provided the data.

Table~1  gives the  distribution  of  the citations  of  the 200  HIPs
published in  1998 as a function  of the most  highly cited facilities
(we  display here  only the  12  most cited  ones), from  HST to  ISO.
Table~2, 3, 4 give the same distribution for the years 1999, 2000, and
2001, respectively.  These tables  show that HST publications have the
highest impact  for the years 1998,  1999, and 2000,  with some strong
challenges from  Keck, Scuba, and  Boomerang, while the impact  of new
space observatories,  like Chandra and XMM-Newton,  is clearly visible
in 2001.

Our  results,  although based  on  a  different  sample of  HIPs,  are
consistent  with the  values found  by Benn  \& S\'anchez  (2001): HST
generates 11\% of the total citations in the years 1995 to 2001.

\begin{figure}[!h]
\plotone {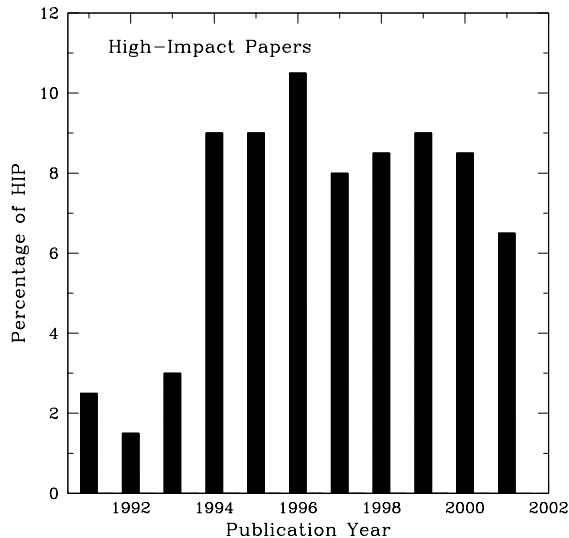}
\caption{Percentage  of  HST  High  Impact  Papers  (HIPs),  from  the
ADS. Since  1994, Hubble has  consistently generated about 8\%  of all
observational HIPs. }
\end{figure}

\section{Top ten most productive HST programs}

Since all 4,116 HST refereed papers are linked in our databases to the
programs having generated the  data, it is straightforward to identify
the most  productive programs.  Table~5 lists the  ten most productive
HST  programs,  in decreasing  numbers  of  related papers.   Column~1
provides  the program type,  Column~2 the  Program I.D.,  Column~3 the
P.I.   last name,  Column~4 the  title  of the  program, Column~5  the
number of papers generated, and Column~6 the total number of citations
to these papers.

It is  worth noting  that all program  categories are present  in this
table,  with  not  only  General  Observer  (GO)  programs,  but  also
Guaranteed  Time  Observer (GTO)  programs,  Parallel (PAR)  programs,
snapshot  (SNAP)  programs,   and  Director  Discretionary  time  (DD)
programs.  This  illustrates the usefulness  of each of  these program
categories.   Not surprisingly,  the top  program is  the  Hubble Deep
Field,  with Williams  as the  PI, which  may have  produced  the most
original, broad impact, and far reaching HST scientific results yet.

\section{Top ten most cited refereed HST publications}

The same databases allow the identification of the most cited refereed
articles based on  Hubble data.  Table~6 lists the  ten most cited HST
papers.  Column~1 gives  the name  of the  first author,  Column~2 the
title of  the paper, and  Column~3 the bibliographic reference  of the
paper.  These papers are  sorted by  decreasing numbers  of citations,
Column~4.

This  list  of  publications   contains  only  papers  presenting  new
scientific  results;  we  have not  included  instrumental/calibration
papers.

\section {Conclusion}

The creation  of effective links  between the STScI and  ADS databases
containing  information  about  all  HST programs,  all  HST  refereed
papers, and their citations, provides us with a powerful and versatile
way to  obtain metrics about describing  the efficiency, productivity,
and scientific impact of the Hubble Space Telescope project.

This may certainly help the funding agencies and the various governing
committees  in shaping  the  future of  HST  time allocation  through
educated  decisions.  The  evaluation of  the present  performances of
space facilities like  HST, Chandra and Spitzer will  help to maximize
the  efficiency and  scientific output  of future  projects,  like the
James Webb Space Telescope (JWST).


\acknowledgments

This research  has made  extensive use of  the NASA  Astrophysics Data
System Bibliographic  services.  We thank  the ADS team  and specially
Michael Kurtz for providing us  with numerous data about citations and
reads of  HST papers.  We are  grateful to the  STScI Librarian, Sarah
Stevens-Rayburn, for  her invaluable input.   Many thanks to  the MAST
team  at STScI,  particularly Karen  Levay, Paolo  Padovani,  and Sara
Anderson for storing and handling the  data used in this study. We are
also grateful to Tim de Zeeuw and Don York for useful comments.


\clearpage

\begin{center}
\begin{tabular}{lc}
\multicolumn{2}{c}{\sc Table 1}\\
\noalign{\smallskip}
\multicolumn{2}{c}{\sc ADS High-Impact Papers 1998}\\
\\
\hline\hline
\noalign{\smallskip}
\multicolumn{1}{c}{Telescope} &Fraction\\
&of the Total\\
\noalign{\smallskip}
\hline
\noalign{\smallskip}
HST        & \llap{1}3.5  \\
Keck       & 7.5\\
Kamiokande & 6.8\\
COBE       & 6.8\\
NOAO       & 7.1\\
ROSAT      & 5.3\\
SCUBA/JCMT & 4.7\\
ASCA       & 4.0\\
Hipparcos  & 3.3\\
ESO        & 2.7\\
\noalign{\smallskip}
\hline
\end{tabular}
\bigskip
\end{center}


\begin{center}
\begin{tabular}{lc}
\multicolumn{2}{c}{\sc Table 2}\\
\noalign{\smallskip}
\multicolumn{2}{c}{\sc ADS High-Impact Papers 1999}\\
\\
\hline\hline
\noalign{\smallskip}
\multicolumn{1}{c}{Telescope} &Fraction\\
&of the Total\\
\noalign{\smallskip}
\hline
\noalign{\smallskip}
HST        & \llap{1}1.8\\
Keck       &  7.6\\
ROSAT      &  7.3\\
SCUBA/JCMT &  5.3\\
Kamiokande &  5.1\\
WHT        &  3.2\\
NOAO       &  3.1\\
ISO        &  2.8\\
ASCA       &  2.5\\
CGRO       &  2.4\\
\noalign{\smallskip}
\hline
\end{tabular}
\bigskip
\end{center}


\begin{center}
\begin{tabular}{lc}
\multicolumn{2}{c}{\sc Table 3}\\
\noalign{\smallskip}
\multicolumn{2}{c}{\sc ADS High-Impact Papers 2000}\\
\\
\hline\hline
\noalign{\smallskip}
\multicolumn{1}{c}{Telescope} &Fraction\\
&of the Total\\
\noalign{\smallskip}
\hline
\noalign{\smallskip}
HST      & \llap{1}2.6\\
Keck     & \llap{1}1.5\\
Chandra  & 7.7\\
Boomerang& 5.8\\
ASCA     & 4.6\\
ESO      & 4.1\\
MAXIMA   & 3.8\\
ISO      & 3.4\\
ROSAT    & 3.4\\
FUSE     & 3.4\\
\noalign{\smallskip}
\hline
\end{tabular}
\bigskip
\end{center}


\begin{center}
\begin{tabular}{lc}
\multicolumn{2}{c}{\sc Table 4}\\
\noalign{\smallskip}
\multicolumn{2}{c}{\sc ADS High-Impact Papers 2001}\\
\\
\hline\hline
\noalign{\smallskip}
\multicolumn{1}{c}{Telescope} &Fraction\\
&of the Total\\
\noalign{\smallskip}
\hline
\noalign{\smallskip}
Chandra     & \llap{1}2.6  \\
XMM-Newton  & \llap{1}1.9  \\
Keck        & 9.6   \\
HST         & 8.9   \\
ESO         & 7.8   \\
AAT         & 4.9   \\
MAXIMA      & 4.0   \\
NOAO        & 3.9   \\
SDSS        & 3.6   \\
ROSAT       & 2.3   \\
\noalign{\smallskip}
\hline
\end{tabular}
\bigskip
\end{center}


\begin{deluxetable}{lcllcc}
\tabletypesize{\small}
\tablenum{5}
\tablecaption{Top ten most productive programs\label{tbl-5}}
\tablewidth{0pt}
\tablehead{
\colhead{Program Type}& \colhead{Program I.D.\tablenotemark{1}}& \colhead{PI}  & \colhead{Title}   &
\colhead{Papers}  & \colhead{Citations}}

\startdata

GO/DD  & 6337 & Williams & Hubble Deep Field & \llap{1}19&5232\\
GO/PAR & 5369 & Griffiths& Medium Deep Survey& 88&2029\\
GO     & 2424 & Bahcall  & Quasar Absorption Line Survey& 58&1953 \\
GO     & 2227 & Mould    & Determination of the Extragalactic Distance Scale  
                    & 57 & 2469\\
SNAP   & 5476 & Sparks   & 3CR Radio Galaxies&57&\phn719\\
GO/DD  & 8058 & Williams & Hubble Deep Field South& 48&\phn772\\
GTO    & 5236 & Westphal & Nuclei of Nearly Normal Galaxies& 48&1973\\
SNAP   & 5479 & Malkan   & Subarcsecond Structure of AGN&45&\phn485\\
GO     & 2563 & Kirshner & SINS The Supernova Intense Study&40&\phn524\\
SNAP   & 7330 & Mulchaey & The Fueling of Active Nuclei&40&\phn416\\
 \enddata
\tablenotetext{1}{Large or multicycle programs may acquire different I.D. numbers
when scheduled through more than one cycle}

\end{deluxetable}


\begin{deluxetable}{lllc}
\tablenum{6}
\tablecaption{Top  ten most  cited HST papers\label{tbl-6}}         
\tablewidth{0pt}          
\tablehead{\colhead{First author}&\colhead{Title}&
           \colhead{Reference}& \colhead{Citations}}

\startdata

Madau      & High redshift Galaxies in the Hubble Deep Field &  1996, MNRAS, 283, 1388 & 701\\
Williams   & The Hubble Deep Field   & 1996, AJ,112, 1335  			       & 586\\
Magorrian  & The Demography of Massive Dark Objects  &1998, AJ, 115, 2285 	       & 568\\
Perlmutter & Discovery of a supernova explosion  &1998, Nature, 391, 51  	       & 397\\
Gebhardt   & Black Hole Mass and Galaxy Velocity Dispersion  & 2000, ApJ, 539, L13     & 378\\
Freedman   & Results from the HST key project to measure H$_{0}$ & 2001, ApJ, 553, 47  & 370\\
Freedman   & Distance to the Virgo Cluster Galaxy M100 &1994, Nature, 371, 757	       & 301\\
Stetson    & The center of the core-cusp globular cluster  &1994, PASP, 106, 250       & 278\\
Freedman   & The HST Extragalactic Distance Scale  &1994, ApJ, 427, 628		       & 271\\
Lowenthal  & Keck Spectroscopy of  z$\sim$3 Galaxies in the HDF & 1997, ApJ, 481, 673L      & 271\\ 
 \enddata

\end{deluxetable}

\end{document}